\documentclass[dvips,11pt]{article}
\usepackage{varioref,exscale,latexsym,amsmath,amssymb}
\usepackage{graphicx}
\usepackage{feynmp} %{feynmf}
\def\beq{\begin{equation}}

\def\eeq{\end{equation}}

\def\beqa{\begin{eqnarray}}

\def\eeqa{\end{eqnarray}}

\title{{\bf Sigma exchange in the nuclear force and effective field theory}}
\author{John F. Donoghue, \\ \\
Department of Physics\\
University of Massachusetts\\
Amherst, MA  01003, USA\\
and \\
Institut des Hautes \'{E}tudes Scientifiques \\
Bures sur Yvette, F-91440, France
 \\}
\begin{document}
\begin{titlepage}
\maketitle
\begin{abstract}
 In the
phenomenological description of the nuclear interaction a crucial
role is traditionally played by the exchange of a scalar I=0 meson,
the sigma, of mass 500-600 MeV, which however is not seen clearly in
the particle spectrum and which has a very ambiguous status in QCD.
I show that a remarkably simple and reasonably controlled
combination of ingredients can reproduce the features of this part
of the nuclear force. The use of chiral perturbation theory
calculations for two pion exchange supplemented by the Omnes
function for pion rescattering suffices to reproduce the magnitude
and shape of the exchange of a supposed $\sigma$ particle, even
though no such particle is present in this calculation. I also show
how these ingredients can describe the contact interaction that
enters more modern descriptions of the internucleon interaction.
\end{abstract}
\vspace{0.2 in}
\end{titlepage}

When describing QCD to non-physicists, we generally say that it is
the theory that accounts for nuclear binding. However, in practice
our understanding of the precise way that QCD leads to nuclear bound
states is still not good. Nuclear binding is traditionally described
by an internucleon potential which can be parameterized by the
exchange of mesons\cite{brown, ericson}. The most important exchange
producing binding in the central potential is a scalar isoscalar
meson, the sigma, of mass around 500-600 MeV. While other exchanges
in the potential are correlated with clear resonances seen in the
particle spectrum, the sigma is a puzzle. It is not seen in the
usual way in the spectrum and, after 40 years of debate, does not
have a clear interpretation in terms of the quarks and gluons of
QCD\footnote{A discussion of the status of the sigma which is very
much in the spirit of the present work can be found in
\cite{Meissner:1990kz}. A careful recent analysis of $\pi\pi$
scattering describing the sigma as a pole on second sheet, quite far
from the real axis, is found in \cite{ccl}}. It is unfortunate that
the key ingredient in the signature effect of the strong
interactions has such an ambiguous status.

The expectation is that the sigma represents, in some way, the
exchange of two pions. The quantum numbers certainly are correct for
this. Sophisticated attempts that construct the potential from
scattering data (e.g. \cite{vinhmau}) have two pions as the lightest
intermediate state. However, while phenomenologically useful, these
are not able to answer the question of the fundamental nature of the
sigma effect. Modern descriptions of the internucleon interaction
use chiral perturbation theory to calculate two pion exchange at low
energy\cite{eft}. However, these do not produce the sigma effect
because the chiral amplitudes grow monotonically with the energy. In
this paper I add a simple and well-motivated addition to the chiral
description, i.e. the Omnes function describing pion rescattering.
We will see that this will produce an interaction remarkably close
in structure to the exchange of a 600 MeV sigma meson. It is clear
that there is no resonance in this description, yet the needed
properties of sigma exchange are reproduced\footnote{Other attempts
to describe the nuclear interaction without a sigma are seen in
\cite{others}}.

Recently, there have been successful applications of ideas of
effective field theory in which the nuclear interaction is treated
not by potentials but by contact interactions - delta function
interactions\cite{eft, serot}. At low energy (recall that the energy
typical of nuclear binding is 10 MeV/nucleon) the result of the
exchange of a heavy particle can be described by a local
interaction. Mathematically, this is consistent with the potential
description because, as the mass $m$ gets large, the Yukawa
potential forms a representation of a delta function. Physically,
this follows from the uncertainty principle, as the exchange of a
heavy particle has a short range. Nonlocality, to the extent it is
needed, can then be described by contact derivative interactions.
This development greatly increases the generality of the description
of nuclei, as it reduces multiple potentials with different
functional forms to a small number of constants giving the strengths
of the contact interactions. I will also show how the use of chiral
perturbation theory plus the Omnes function can provide a reasonably
controlled calculation of the I = 0 scalar contact interaction.

In QCD, the low energy interactions of two nucleons obeys an
unsubtracted dispersion relation in the different partial
waves\cite{brown, ericson, vinhmau} Specializing to the scalar
isoscalar channel, we can write the momentum space and coordinate
space interaction as
\begin{eqnarray}
{V}_S(q^2) &=&  \frac{2}{\pi}\int_{2m_\pi}^\infty d\mu~\mu
\frac{\rho_S(\mu)}{\mu^2+q^2}  \nonumber \\
V_S(r) &=&  \frac{1}{2 \pi^2 r}\int_{2m_\pi}^\infty d\mu~\mu
~e^{-\mu r}~{\rho}_S(\mu)
\end{eqnarray}
Here $\rho_S$ describes the physical intermediate states that occur
at energy $\mu$ in the crossed channel. The threshold occurs at
twice the pion mass and the lowest energy intermediate state is two
pions.

At low energy these spectral functions can be rigorously calculated
in chiral perturbation theory. The imaginary part of the Feynman
diagrams describe the physical intermediate states and generate the
spectral function $\rho_S$.
\begin{figure}[h]
 \begin{fmffile}{fmftwopion}
  \begin{equation*}
   \begin{fmfgraph*}(100,50)
    \fmfleft{vleftlow,vlefthigh} %set one vertex on the left
    \fmfright{vrightlow,vrighthigh} %set one vertex on the right
    \fmf{plain}{vleftlow,vmiddlelow1,vmiddlelow2,vmiddlelow3,vrightlow} %draw a plain line between the vertices
        \fmf{plain}{vlefthigh,vmiddlehigh1,vmiddlehigh2,vmiddlehigh3,vrighthigh} %draw a plain line between the vertices
    \fmfdot{vmiddlelow1,vmiddlelow3,vmiddlehigh2}
\fmfv{label=$N$,label.angle=180}{vlefthigh}
\fmfv{label=$N$,label.angle=180}{vleftlow}
\fmfv{label=$N$,label.angle=0}{vrighthigh}
\fmfv{label=$N$,label.angle=0}{vrightlow}
    \fmffreeze
    \fmf{dashes,label=$\pi$,label.side=left}{vmiddlelow1,vmiddlehigh2}
        \fmf{dashes,label=$\pi$,label.side=right}{vmiddlelow3,vmiddlehigh2}
        \fmfv{label={(a)},label.angle=-90,label.dist=15}{vmiddlelow2}
   \end{fmfgraph*}  {}\hspace*{30pt}
      \begin{fmfgraph*}(100,50)
    \fmfleft{vleftlow,vlefthigh} %set one vertex on the left
    \fmfright{vrightlow,vrighthigh} %set one vertex on the right
    \fmf{plain}{vleftlow,vmiddlelow1,vmiddlelow2,vmiddlelow3,vrightlow} %draw a plain line between the vertices
        \fmf{plain}{vlefthigh,vmiddlehigh1,vmiddlehigh2,vmiddlehigh3,vrighthigh} %draw a plain line between the vertices
    \fmfdot{vmiddlelow2,vmiddlehigh3,vmiddlehigh1}
\fmfv{label=$N$,label.angle=180}{vlefthigh}
\fmfv{label=$N$,label.angle=180}{vleftlow}
\fmfv{label=$N$,label.angle=0}{vrighthigh}
\fmfv{label=$N$,label.angle=0}{vrightlow}
    \fmffreeze
    \fmf{dashes,label=$\pi$,label.side=right}{vmiddlehigh1,vmiddlelow2}
        \fmf{dashes,label=$\pi$,label.side=left}{vmiddlehigh3,vmiddlelow2}
        \fmfv{label={(b)},label.angle=-90,label.dist=15}{vmiddlelow2}
   \end{fmfgraph*}  {}\hspace*{30pt}
   \begin{fmfgraph*}(100,50)
    \fmfleft{vleftlow,vlefthigh} %set one vertex on the left
    \fmfright{vrightlow,vrighthigh} %set one vertex on the right
    \fmf{plain}{vleftlow,vmiddlelow1,vmiddlelow2,vmiddlelow3,vrightlow} %draw a plain line between the vertices
        \fmf{plain}{vlefthigh,vmiddlehigh1,vmiddlehigh2,vmiddlehigh3,vrighthigh} %draw a plain line between the vertices
    \fmfdot{vmiddlelow2,vmiddlehigh2}
\fmfv{label=$N$,label.angle=180}{vlefthigh}
\fmfv{label=$N$,label.angle=180}{vleftlow}
\fmfv{label=$N$,label.angle=0}{vrighthigh}
\fmfv{label=$N$,label.angle=0}{vrightlow}
    \fmffreeze
  \fmf{dashes,left=0.5,label=$\pi$,label.side=left}{vmiddlelow2,vmiddlehigh2}
    \fmf{dashes,right=0.5,label=$\pi$,label.side=right}{vmiddlelow2,vmiddlehigh2}
      \fmfv{label={(c)},label.angle=-90,label.dist=15}{vmiddlelow2}
   \end{fmfgraph*}  {}\hspace*{30pt}
  \end{equation*}
  \vspace{0.5in}
    \begin{equation*}
   \begin{fmfgraph*}(100,50)
    \fmfleft{vleftlow,vlefthigh} %set one vertex on the left
    \fmfright{vrightlow,vrighthigh} %set one vertex on the right
    \fmf{plain}{vleftlow,vmiddlelow1,vmiddlelow2,vmiddlelow3,vrightlow} %draw a plain line between the vertices
        \fmf{plain}{vlefthigh,vmiddlehigh1,vmiddlehigh2,vmiddlehigh3,vrighthigh} %draw a plain line between the vertices
    \fmfdot{vmiddlelow1,vmiddlelow3,vmiddlehigh1,vmiddlehigh3}
\fmfv{label=$N$,label.angle=180}{vlefthigh}
\fmfv{label=$N$,label.angle=180}{vleftlow}
\fmfv{label=$N$,label.angle=0}{vrighthigh}
\fmfv{label=$N$,label.angle=0}{vrightlow}
    \fmffreeze
    \fmf{dashes,label=$\pi$,label.side=left}{vmiddlelow1,vmiddlehigh1}
        \fmf{dashes,label=$\pi$,label.side=right}{vmiddlelow3,vmiddlehigh3}
        \fmfv{label={(d)},label.angle=-90,label.dist=15}{vmiddlelow2}
   \end{fmfgraph*}  {}\hspace*{30pt}
      \begin{fmfgraph*}(100,50)
    \fmfleft{vleftlow,vlefthigh} %set one vertex on the left
    \fmfright{vrightlow,vrighthigh} %set one vertex on the right
    \fmf{plain}{vleftlow,vmiddlelow1,vmiddlelow2,vmiddlelow3,vrightlow} %draw a plain line between the vertices
        \fmf{plain}{vlefthigh,vmiddlehigh1,vmiddlehigh2,vmiddlehigh3,vrighthigh} %draw a plain line between the vertices
    \fmfdot{vmiddlelow1,vmiddlelow3,vmiddlehigh3,vmiddlehigh1}
\fmfv{label=$N$,label.angle=180}{vlefthigh}
\fmfv{label=$N$,label.angle=180}{vleftlow}
\fmfv{label=$N$,label.angle=0}{vrighthigh}
\fmfv{label=$N$,label.angle=0}{vrightlow}
    \fmffreeze
    \fmf{dashes}{vmiddlehigh1,vmiddlelow3}
        \fmf{dashes}{vmiddlehigh3,vmiddlelow1}
\fmfv{label=$\pi$,label.side=left,label.dist=20,label.angle=-80}{vmiddlehigh1}
\fmfv{label=$\pi$,label.side=left,label.dist=20,label.angle=-100}{vmiddlehigh3}
        \fmfv{label={(e)},label.angle=-90,label.dist=15}{vmiddlelow2}
   \end{fmfgraph*}  {}\hspace*{30pt}
  \end{equation*}
 \end{fmffile}
$\hspace*{0.515\textwidth}$ \vspace{0 in} \caption{Two pion exchange
diagrams which arise in chiral perturbation theory.} \vspace*{10pt}
\end{figure}
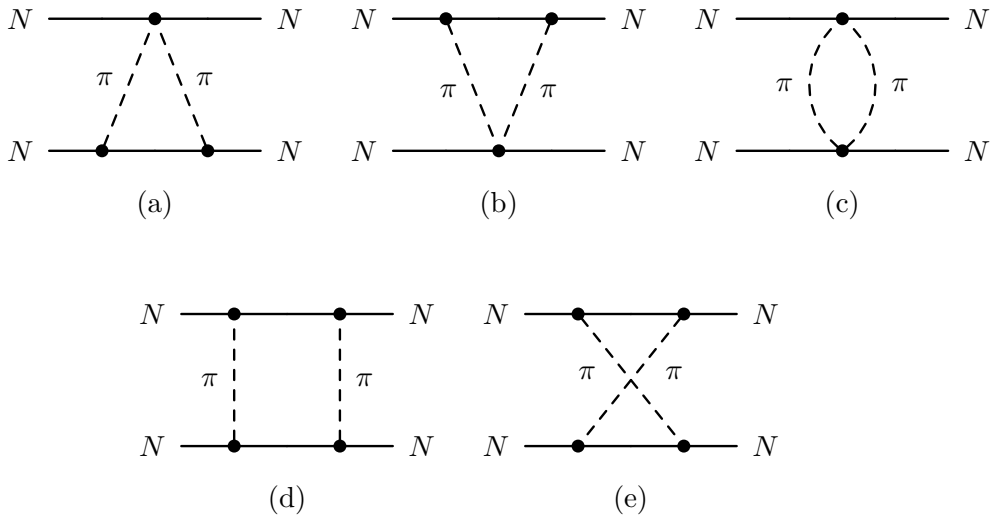
For the diagrams of Figs 1 a,b,c these imaginary parts
are\cite{NNLO, NNNLO}
\begin{equation}
 \rho_S^{\rm a,b} (\mu) = \frac{3 g_A^2}{64
F_\pi^4}\left[4c_1 m_\pi^2 +c_3(\mu^2-2m_\pi^2)
\right]\frac{(\mu^2-2m_\pi^2)}{\mu}~\theta (\mu -2m_\pi)
\end{equation}
\begin{eqnarray}
\rho_S^{\rm c} (\mu) &=& -\frac{3}{32 \pi
F_\pi^4}\sqrt{1-\frac{4m_\pi^2}{\mu^2}} ~\theta (\mu -2m_\pi)  \\
& &~~~ \left(\left[4c_1 m_\pi^2 +\frac{c_2}{6}(\mu^2 - 4m_\pi^2)
+c_3(\mu^2-2m_\pi^2) \right]^2 +\frac{c_2^2}{45}(\mu^2-4
m_\pi^2)^2\right)\nonumber
\end{eqnarray}
Here $c_1, c_2, c_3$ are parameters that describe the $NN\pi\pi$
vertex - these have been measured in pion nucleon
interactions\cite{NNLO, NNNLO, lattice}. I will address the box and
crossed box diagrams below. These spectral functions are valid in
the low energy regime only, and one observes that they grow
monotonically with the energy.

However, there is another ingredient which necessarily  enters. In
the description of the $\pi\pi$ system, unitarity requires the
inclusion of $\pi\pi$ rescattering. For a single elastic partial
wave, unitarity of the S matrix and analyticity require a unique
form of the solution, given originally by Omnes\cite{Omnes}. The
amplitudes in the elastic region are described by a polynomial in
the energy times the Omnes function
\begin{equation}
\Omega (\mu) = exp[\frac{\mu^2}{\pi} \int
\frac{ds}{s}~\frac{\delta(s)}{s-\mu^2}]
\end{equation}
Here $\delta$ is the $\pi\pi$ scattering phase shift, in our case
for the I=0, J=0 channel. Chiral perturbation theory is consistent
with this order by order in the energy expansion. Following Ref.
\cite{dgl}, it is known how to match this general description to the
results of chiral perturbation theory by appropriately identifying
the polynomial. The elastic region in this channel extends
effectively up to energies of 1000 MeV.

In practice there has been good success at using the lowest order
chiral amplitudes, supplemented by the Omnes function. An example
close to the present problem is $\gamma\gamma\to \pi\pi$ in the S
wave. Here a lowest order calculation supplemented with an Omnes
function\cite{dhl} yields results in close agreement with both
experiment and with a two loop chiral calculation up to energies
beyond 700 MeV\cite{twoloop}.

I will adopt the Omnes solution matched to the leading order chiral
result, and will explore possible modifications below. The
description of the spectral function then becomes
\begin{equation}
\rho_S(\mu) = \rho_S^{a,b} Re \Omega(\mu) +
\rho_S^{c}|\Omega(\mu)|^2
\end{equation}
The phase shifts can be analyzed in chiral perturbation theory in
combination with experiment, with the definitive treatment of
Colangelo et al (CGL)\cite{cgl}. Their result for the I = 0, J=0
phase shift is shown in Fig 2, along with the resulting Omnes
functions. Note that there is no sigma resonance in the phase shift
near 300 - 600 MeV. A resonance in the elastic region is manifest by
the phase shift passing through 90 degrees, which certainly does not
happen near the sigma mass. (If one explores the complex plane there
is a pole on the second sheet very far from the real
axis\cite{ccl}.) In producing the Omnes function, I had to extend
the phase shifts above the $\mu$ = 850 MeV endpoint of the CGL
analysis in order that the principle value part of the Omnes
function integral be well behaved near the upper end. As long as
this extension is smooth it has little effect on this calculation.

\begin{figure}
\begin{center}
 \begin{minipage}[t]{.07\textwidth}
    \vspace{0pt}
%    \centering
 %   \vspace*{120pt}
    \hspace*{-180pt}
    \rotatebox{0}{$\delta$}
    \hspace*{320pt}{$|\Omega|^2$} \\ \\ \\ \\
    \hspace*{120pt}{$Re\Omega$}
     \end{minipage}
  \begin{minipage}[t]{0.93\textwidth}
  \vspace{-90pt}
$\begin{array} {c@{\hspace{0.01 in}}c} \multicolumn{1}{l}{} &
\multicolumn{1}{l}{} \\
{\resizebox{2.5in}{!}{\includegraphics{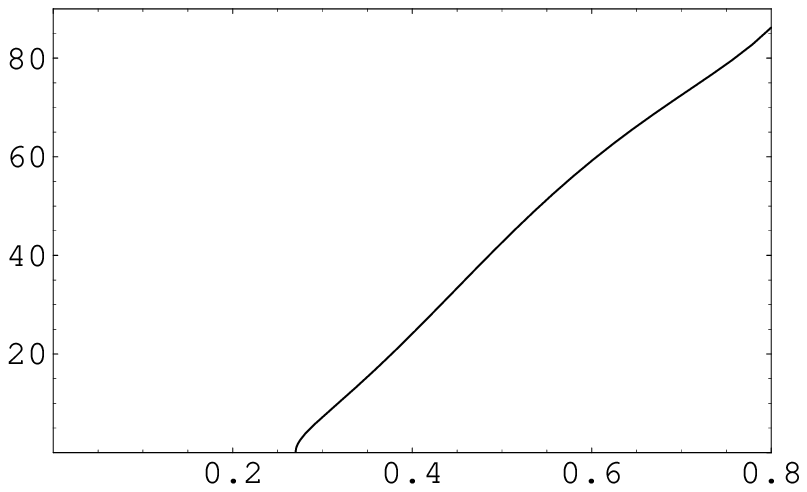}}}&
{\resizebox{2.5in}{!}{\includegraphics{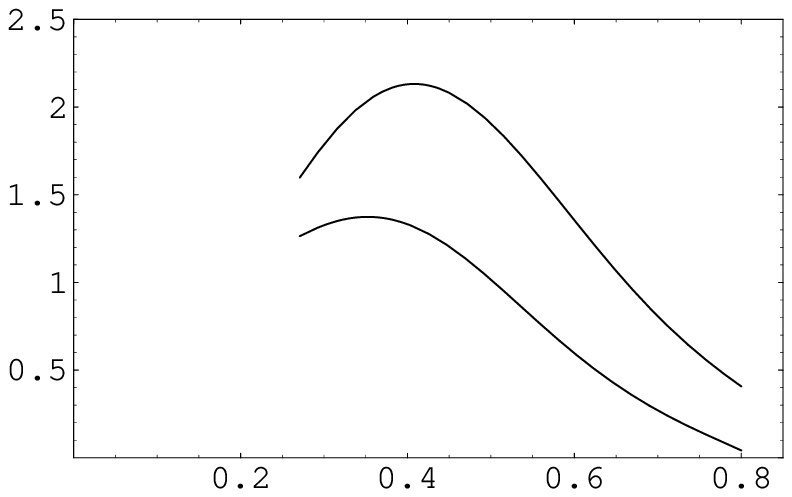}}} \\
\mbox{\hspace*{100pt} {$\mu$} ~(GeV)} & \mbox{\hspace*{75pt} {$\mu$}~(GeV)} \\
\mbox{\hspace*{25pt} (a)} & \mbox{\hspace*{25pt} (b)}
\end{array}$
  \end{minipage}

\end{center}
\caption{The left graph gives the pion phaseshifts that are the
input into the Onmes function, while the right figure shows the real
part and the absolute square of the Omnes function.} \label{run}
\end{figure}

With these ingredients, we can display the result for the scalar
interaction. In Fig. 3, I show the result for $\rho$, along with the
individual contributions of the diagrams of Fig 1. If we had a pure
sigma exchange this would be delta function at the mass of the
$\sigma$, or a Breit-Wigner shape corresponding to a narrow
resonance. One could be forgiven for seeing this result as a very
broad resonance, even though no resonance exists in the formalism.
The coordinate space potential is shown in Fig. 3. Also shown for
comparison is the potential of an infinitely narrow 600 MeV scalar
with a normalization chosen to match. In practice these are hard to
differentiate because the curves are nearly identical. The simple
description of Eq. 5 reproduces closely the spatial variation of the
sigma potential. The strength of the interaction will be addressed
below.

\begin{figure}
\begin{center}
 \begin{minipage}[t]{.07\textwidth}
    \vspace{0pt}
%    \centering
 %   \vspace*{120pt}
    \hspace*{-193pt}
        \hspace*{300pt} {$r$~(GeV$^{-1}$)}\\ \\
%    \vspace*{60 pt}
    \hspace*{12pt}{$rV(r)$}
\hspace*{-205pt} \rotatebox{0}{${\rho}/{\mu}$}
 %   \hspace*{175pt}{$rV(r)$} \\
     \hspace*{130pt}{a,b}\\
    \hspace*{-25pt}{c} \\ \\
    \hspace*{-60pt}{total}
     \end{minipage}
  \begin{minipage}[t]{0.93\textwidth}
  \vspace{-90pt}
$\begin{array} {c@{\hspace{0.01 in}}c} \multicolumn{1}{l}{} &
\multicolumn{1}{l}{} \\
{\hspace{-30pt}\resizebox{2.5in}{!}{\includegraphics{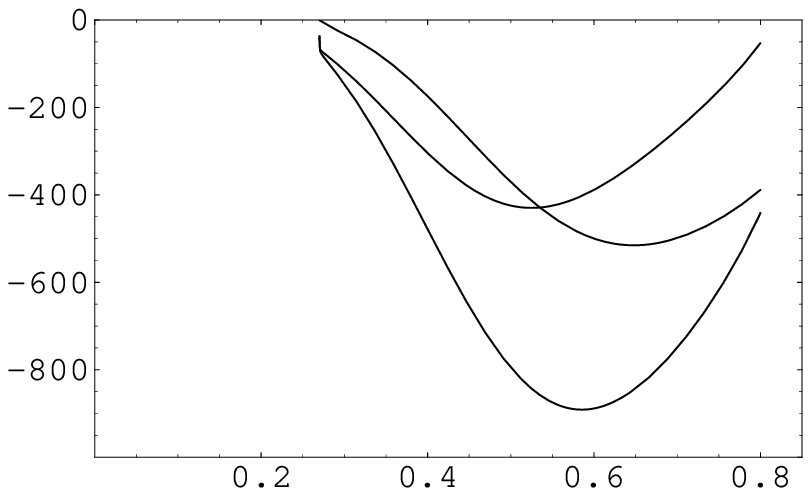}}}&
~~~~~~~~{\resizebox{2.5in}{!}{\includegraphics{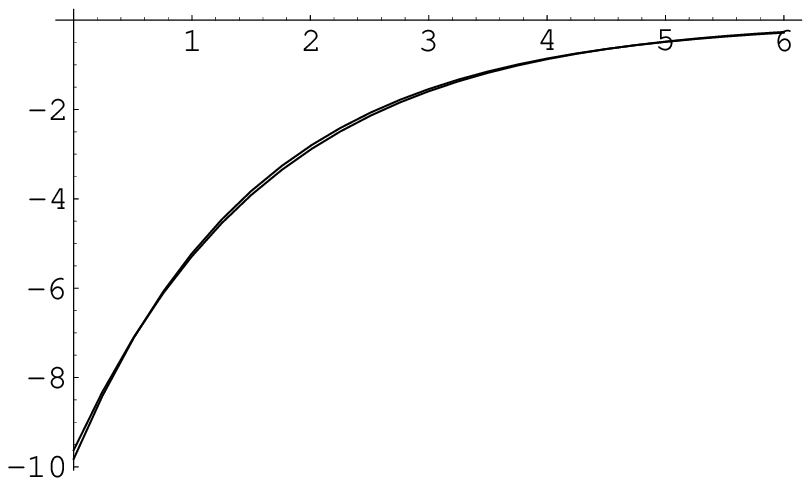}}} \\
\mbox{\hspace*{100pt} {$\mu$} ~(GeV)} & \mbox{\hspace*{75pt} }\\
\mbox{\hspace*{25pt} (a)} & \mbox{\hspace*{25pt} (b)}
\end{array}$
  \end{minipage}
\end{center}
\caption{The left figure shows the our results for the spectral
function $\rho(\mu)/\mu$ as well as the individual components of
diagram 1 a,b,c. The right figure shows the coordinate space
potential $rV(r)$. There are actually two curves in the figure on
the right. One is the result of this calculation and the second is
that of a narrow 600~MeV sigma, with normalization chosen to match.
The curves cannot be distinguished. } \label{results}
\end{figure}

One can address the robustness of this result by considering
possible higher order modifications of the basic representation. The
$NN\pi\pi$ interaction has been described by the lowest order chiral
Lagrangian. There are also energy dependent modifications to these
low order results. In particular, we expect that there might be form
factors depending on the energy. To probe this effect, let as modify
the $\pi\pi$ interaction by a form factor.
\begin{equation}
c_3 \to \frac{c_3}{(\mu^2+m^2)^n}
\end{equation}
The choice of $n=1$ and $m=800$ is shown in Fig. 4. While the
relative contribution of the two diagrams change (since the latter
has the form factor squared) and the magnitude is different, the
energy variation and spatial variation are remarkably similar to the
original case. Use of a dipole form factor does not change this
conclusion.
\begin{figure}[h]
%\vspace*{5pt}
\begin{center}
  \begin{minipage}[t]{.07\textwidth}
    \vspace{0pt}
    \centering
    \vspace*{0 in}
    \hspace*{70pt}{$\rho(\mu)/\mu$} \\
         \hspace*{250pt}{a,b} \\
    \hspace*{265pt}{c} \\
    \hspace*{0pt} \\
    \hspace*{245pt}{tot}
  \end{minipage}%
  \begin{minipage}[t]{0.93\textwidth}
    \vspace{0 in}
    \centering
\hspace{-0.0 in}\rotatebox{-0}
{\includegraphics[width=0.6\textwidth,height=!]{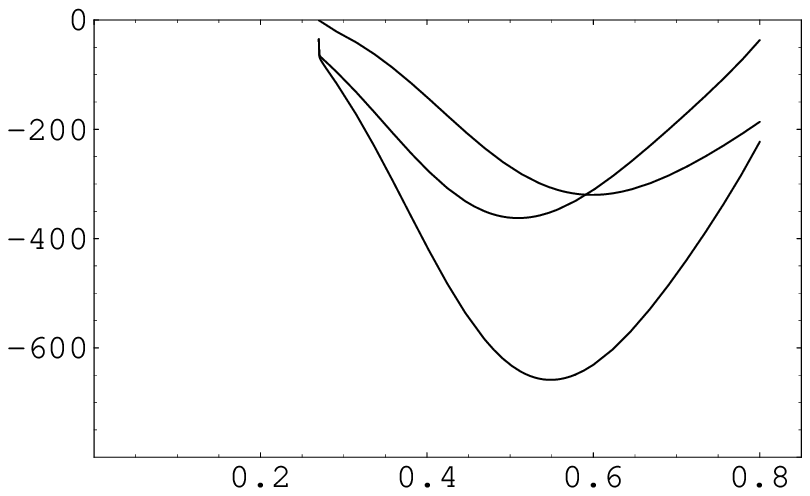}}\\
    ${} \vspace*{-0pt}$
\mbox{\hspace*{100pt} {$\mu$} ~(GeV)}\\
  \end{minipage}
\end{center}
$\hspace*{0.515\textwidth}$ \vspace{0 in} \caption{The spectral
function  $\rho(\mu)/\mu$ in the presence of higher order energy
dependent modifications as described in the text } \vspace*{10pt}
\end{figure}

We can best address the strength of the interaction by describing
the magnitude of the contact interaction $G_s$. For a narrow sigma
this would have the value $g_\sigma^2/m_\sigma^2$. There is some
uncertainty in the appropriate value of $G_S$, because the preferred
fit magnitudes depend somewhat on the calculational scheme used.
However, phenomenological studies of the nuclei tend to require $G_S
= 300-450$~MeV$^{-2}$ . The contact interaction is given by the
strength of the momentum space potential of Eq. 1 evaluated at
$q^2=0$. This is just integral under the integrand shown in Fig. 3.
The result depend most sensitively on the parameter $c_3$, which is
not perfectly known. The phenomenological extraction of $c_3$ from
$\pi N$ data has a large error bar, $c_3 =
-4.7^{+1.2}_{-1.0}$~GeV$^{-2}$~ \cite{NNLO, lattice}. However, when
using an Omnes representation, it is likely that this constraint is
on the product $c_3\Omega(2m_\pi)$, in which case the value would be
$c_3 = -3.7^{+1.0}_{-0.8}$~GeV$^{-2}$. (The other parameter choices
used were $c_1=-0.64 $~GeV$^{-2}$ and $c_2=3.3 $~GeV$^{-2}$,
although these have only a small impact on the results.) The result
for $G_S$ as a function of $c_3$ is shown in Fig 4. There is good
agreement for the required range of magnitudes of $G_S$ for the
allowed values of $c_3$\cite{eft}. Here the use of a form factor
does make a difference. With the form factor described above the
value of $G_S$ is 40\% smaller than without it for a given value of
$c_3$. However at present understanding this difference may be
accounted for by adjusting the value of $c_3$. These uncertainties
in the appropriate values of $c_3$ and $G_S$ keep us from using the
magnitude as a precise test of the method.
\begin{figure}[h]
%\vspace*{5pt}
\begin{center}
  \begin{minipage}[t]{.07\textwidth}
    \vspace{0pt}
    \centering
    \vspace*{0 in}
    \hspace*{70pt}{$G_S$}
  \end{minipage}%
  \begin{minipage}[t]{0.93\textwidth}
    \vspace{0 in}
    \centering
\hspace{-0.0 in}\rotatebox{-0}
{\includegraphics[width=0.6\textwidth,height=!]{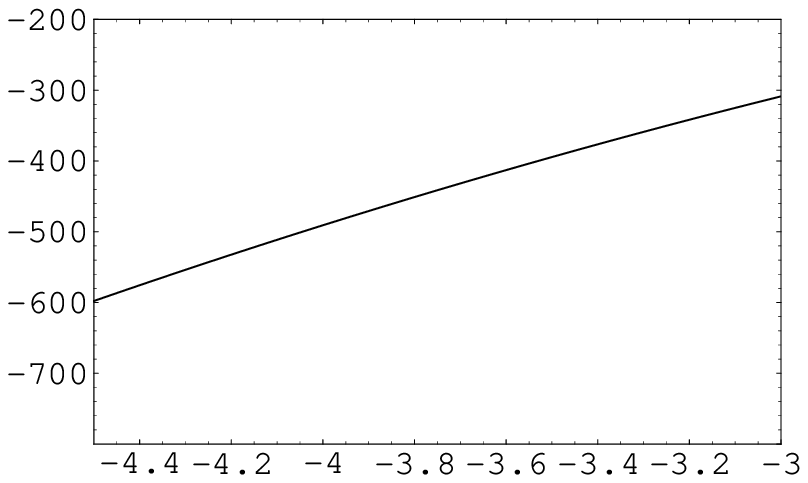}}\\
    ${} \vspace*{-0pt}$
    \mbox{\hspace*{100pt} {$c_3$} ~(GeV$^{-2}$)} \\
  \end{minipage}
\end{center}
$\hspace*{0.515\textwidth}$ \vspace{0 in} \caption{The strength of
the scalar interaction as a function of the parameter $c_3$. }
\vspace*{10pt}
\end{figure}

In an effective field theory treatment, one keeps pion exchange as
an explicit degree of freedom while treating the shorter range
interactions as contact terms. In such a treatment, we should treat
the box and crossed box diagrams of Fig 1 dynamically, and they
should therefore not be included into the contact interaction. For
this reason I did not include these diagrams in the calculation of
the integrand $\rho_S(\mu)/\mu$ whose integral gives the strength of
the contact interaction. However, it should be admitted that the
different approximations schemes that are used to calculate nuclear
properties treat the iteration of the one-pion exchange interaction
in quite different ways. Use of time-dependent perturbation theory,
field theoretic methods, mean-field methods, Bethe-Salpeter
approximations, etc all capture different amounts of the box and
crossed box diagrams. This is related to the serious ambiguity of
what is meant by a potential in field theory\cite{sucher}, and this
likely accounts for the range of fit values of $G_S$. It seems that
for the scalar central potential the iteration of the one pion
interaction is a numerically small compared to the irreducible two
pion/sigma contribution, for example see \cite{NNLO} and Fig 3.15 of
Ref. \cite{ericson}. Hopefully this ambiguity will be cleared up as
effective field theory techniques are extended to heavier
nuclei\cite{summary}.

Chiral perturbation theory plus the Omnes function give a quite
simple description of the scalar central potential, with a result
very similar to the exchange of a fictitious sigma particle. However
fortunately there is no need in such a description to postulate such
a scalar particle. This description appears to be robust, being
qualitatively unchanged by the addition of higher order
interactions. Besides elucidating a long standing puzzle, these
results are useful because we have a reasonably solid control over
all the ingredients, the chiral amplitudes and the $\pi\pi$ phase
shifts. Questions such as the quark mass dependence of nuclear
bending may now be addressed in a reasonable
fashion\cite{chirallimit}. The connection of the nuclear interaction
to QCD becomes more under control.

\section*{Acknowledgement} I would like to thank especially Thibault
Damour for many conversations during the course of this work, U. von
Kolck for discussions about the iterated one pion interaction, R.
Furnstahl and B. Serot for information on the nuclear interaction
and J. Gasser, Barry Holstein and H. Leutwyler for continually
informative discussions on the dispersion representation and the
Omnes function. I also appreciate the hospitably the IHES where most
of this work was performed. This work has been partially supported
by the U.S National Science Foundation.


\begin{thebibliography}{99}

\bibitem{brown}
G.~E.~Brown and A.~D.~Jackson, {\it The Nucleon-Nucleon
Interaction},(North Holland, 1977).

\bibitem{ericson}
T.~E.~O.~Ericson and W.~Weise, {\it Pions And Nuclei},(Clarendon,
Oxford, 1988)
\bibitem{Meissner:1990kz}
  U.~G.~Meissner,
  ``Chiral Dynamics: Where Are The Scalars?,''
  Comments Nucl.\ Part.\ Phys.\  {\bf 20}, 119 (1991).

\bibitem{ccl}
  I.~Caprini, G.~Colangelo and H.~Leutwyler,
  ``Mass and width of the lowest resonance in QCD,''
  arXiv:hep-ph/0512364.

\bibitem{vinhmau}
W.~Cottingham and R.~ Vinh Mau, ``Theoretical nucleon-nucleon
potential'', Phys. Rev. {\bf 130}, 735 (1963).\\
W.~N.~Cottingham, M.~Lacombe, B.~Loiseau, J.~M.~Richard and R.~Vinh
Mau, ``Nucleon Nucleon Interaction From Pion Nucleon Phase Shift
Analysis,'' Phys.\ Rev.\ D {\bf 8}, 800 (1973). \\
M.~Chemtob, J.~W.~Durso and D.~O.~Riska, ``Two-Pion-Exchange
Nucleon-Nucleon Potential,''  Nucl.\ Phys.\ B {\bf 38} (1972) 141.

\bibitem{others}
W.~Grein and P.~Kroll, ``Two Pion And Three Pion Cut Contributions
To Nucleon-Nucleon Scattering,'' Nucl.\ Phys.\ A {\bf 338}, 332
(1980). \\
F.~Wang, G.~h.~Wu, L.~j.~Teng and T.~Goldman, ``Quark
delocalization, color screening, and nuclear intermediate range
attraction,'' Phys.\ Rev.\ Lett.\  {\bf 69}, 2901 (1992)
[arXiv:nucl-th/9210002]. \\
E.~Oset, H.~Toki, M.~Mizobe and T.~T.~Takahashi, ``$\sigma$ exchange
in the N N interaction within the chiral unitary approach,'' Prog.\
Theor.\ Phys.\  {\bf 103}, 351 (2000)
  [arXiv:nucl-th/0011008].


\bibitem{eft}
E.~Epelbaum, W.~Glockle, A.~Kruger and U.~G.~Meissner, ``Effective
theory for the two-nucleon system,'' Nucl.\ Phys.\ A {\bf 645}, 413
(1999) [arXiv:nucl-th/9809084].\\
M.~J.~Savage, ``Effective field theory for nuclear physics,''
arXiv:nucl-th/0301058.\\
P.~F.~Bedaque and U.~van Kolck, ``Effective field theory for
few-nucleon systems,'' Ann.\ Rev.\ Nucl.\ Part.\ Sci.\  {\bf 52},
339 (2002) [arXiv:nucl-th/0203055].


\bibitem{serot}
B.~A Nilolaus, T. Hoch and D.~G. Madland, ``Nuclear ground state
properties in a relativistic point coupling model'', Phys. Rev.
{\bf C46}, 1757 (1992)  \\
 B.~D.~Serot and
J.~D.~Walecka, ``Effective field theory in nuclear
many-body physics,'' arXiv:nucl-th/0010031.\\
 B.~D.~Serot and
J.~D.~Walecka, ``Recent progress in quantum hadrodynamics,'' Int.\
J.\ Mod.\ Phys.\ E {\bf 6}, 515 (1997) [arXiv:nucl-th/9701058].\\
R.~J.~Furnstahl and B.~D.~Serot, ``Parameter Counting in
Relativistic Mean-Field Models,'' Nucl.\ Phys.\ A {\bf 671}, 447
(2000) [arXiv:nucl-th/9911019].\\
 J.~J.~Rusnak and R.~J.~Furnstahl,
``Relativistic point-coupling models as effective theories of
nuclei,'' Nucl.\ Phys.\ A {\bf 627}, 495 (1997)
[arXiv:nucl-th/9708040].\\
B. Machleidt and D. R. Entem, ``Towards a consistent approach to
nuclear structure: EFT of two- and many-body forces''
arXive:nucl-th/0503025

\bibitem{NNLO}
N.~Kaiser, R.~Brockmann and W.~Weise, ``Peripheral nucleon nucleon
phase shifts and chiral symmetry,'' Nucl.\ Phys.\ A {\bf 625}, 758
(1997) [arXiv:nucl-th/9706045].\\
E.~Epelbaum, W.~Glockle and U.~G.~Meissner, ``Improving the
convergence of the chiral expansion for nuclear forces.  I:
Peripheral phases,'' arXiv:nucl-th/0304037.
%%CITATION = NUCL-TH 9706045;%%
\bibitem{NNNLO}
N.~Kaiser, ``Chiral 2pi exchange N N potentials: Two-loop
contributions,'' Phys.\ Rev.\ C {\bf 64}, 057001 (2001)
[arXiv:nucl-th/0107064].\\
%%CITATION = NUCL-TH 0107064;%%
  D.~R.~Entem and R.~Machleidt,
  ``Chiral 2$\pi$ exchange at order four and peripheral N N scattering,''
  Phys.\ Rev.\ C {\bf 66}, 014002 (2002)
  [arXiv:nucl-th/0202039].
\bibitem{lattice}
 U.~G.~Meissner,
  ``Quark mass dependence of baryon properties,''
  PoS {\bf LAT2005}, 009 (2005)
  [arXiv:hep-lat/0509029].

\bibitem{Omnes}
  R.~Omnes,
  ``On the solution of certain singular integral equations of quantum field
  theory,''
  Nuovo Cim.\  {\bf 8}, 316 (1958).

%\cite{Donoghue:1990xh}
\bibitem{dgl}
J.~F.~Donoghue, J.~Gasser and H.~Leutwyler, ``The decay of a light
Higgs boson,'' Nucl.\ Phys.\ B {\bf 343}, 341 (1990).  \\
%%CITATION = NUPHA,B343,341;%


 \bibitem{dhl} J.~F.~Donoghue and B.~R.~Holstein,
``Photon-photon scattering, pion polarizability and chiral
symmetry,'' Phys.\ Rev.\ D {\bf 48}, 137 (1993)
[arXiv:hep-ph/9302203]. \\
D.~Morgan and M.~R.~Pennington,``Is low-energy gamma gamma $\to$ pi0
pi0 predictable?,''
  Phys.\ Lett.\ B {\bf 272}, 134 (1991).



\bibitem{twoloop}
  S.~Bellucci, J.~Gasser and M.~E.~Sainio,
  ``Low-energy photon-photon collisions to two loop order,''
  Nucl.\ Phys.\ B {\bf 423}, 80 (1994)
  [Erratum-ibid.\ B {\bf 431}, 413 (1994)]
  [arXiv:hep-ph/9401206].

%\cite{Colangelo:2001df}
\bibitem{cgl}
G.~Colangelo, J.~Gasser and H.~Leutwyler, ``$\pi \pi$ scattering,''
Nucl.\ Phys.\ B {\bf 603}, 125 (2001) [arXiv:hep-ph/0103088].
%%CITATION = HEP-PH 0103088;%%
\bibitem{sucher}
  J.~Sucher,
  ``The concept of potential in quantum field theory,''
  arXiv:hep-ph/9412388.
  %%CITATION = HEP-PH 9412388;%%


\bibitem{summary}
  R.~Machleidt and D.~R.~Entem,
  ``Towards a consistent approach to nuclear structure: EFT of two- and
  many-body forces,''
  arXiv:nucl-th/0503025.
\bibitem{chirallimit}
J.~F.~Donoghue (in preparation)
\end{thebibliography}
\end{document}